 \documentclass[preprint]{aastex}

\shortauthors{Dickel, Strom, \& Milne}
\shorttitle{Supernova Remnant MSH14-6$\it 3$}

\begin{document}


\title{The Radio Structure of the Supernova Remnant G315.4-2.3
  (MSH~14$-$6$\it 3$)}


\author{John R. Dickel\altaffilmark{1}}
\affil{Astronomy Department, University of Illinois at Urbana-
  Champaign, Urbana IL 61801}
\email{johnd@astro.uiuc.edu}

\author{Richard G. Strom\altaffilmark{2}}
\affil{ASTRON, 7990AA Dwingeloo, The Netherlands}
\email{strom@astron.nl}

\and

\author{D. K. Milne}
\affil{Australia Telescope National Facility, Epping NSW 1710,
  Australia}
\email{dmilne@atnf.csiro.au}


\altaffiltext{1}{visiting astronomer at ASTRON, 7990AA Dwingeloo, The
  Netherlands} 

\altaffiltext{2}{ also at Astronomical Institute, University of
  Amsterdam, The Netherlands}


\begin{abstract}
G315.4$-$2.3 is an extended shell supernova remnant (SNR) with some
characteristics of evolutionarily young remnants and some of older
ones.  To further elucidate some of its characteristics, we present
imaging and polarimetry of this SNR at a frequency of 1.34 GHz with a
resolution of 8 arcsec made with the Australia Telescope Compact Array.

The indicators of youth are: Morphologically, the radio emission
arises in a smooth shell without the fine scale filaments seen in the
optical.  Many of the optical filaments are Balmer dominated.  Where
measurable, the orientation of the magnetic field appears to be radial
with respect to the center of the remnant.  There may have been a
supernova in that region in AD~185.  
 
Indications of older age include:  Particularly in RCW~86, the bright
optical nebula in the southwestern corner of this extended SNR, but
also in other locations there are several filaments with bright [S~II]
emission representative of older shocked filaments in radiative
equilibrium.  If the remnant lies at the kinematical distance of 2.8
kpc, it has a diameter of 37 pc which would be large for a remnant less
than two thousand years old.  

The remnant seems to be expanding inside a cavity outlined by infrared
emission and so it could well be young and large.  Where it is
encountering the walls of the cavity it is slowing rapidly and we
observe the radiative filaments.  RCW~86 itself is encountering a dense
clump of material but may also be the remains of a more compact lump of
ejecta ploughing into the surroundings.
\end{abstract}




\section{Introduction}

The object G315.4$-$2.3 was first cataloged as the radio source
MSH~14$-$6{\it 3\/ }by \citet{m61}.  It was soon recognized as a
supernova remnant (SNR) because of its large angular size --- about 45
arcmin in diameter --- shell structure, radio polarization, and
non-thermal radio spectrum \citep{h67}.  The bright nebula RCW~86
(Rodgers et al.\ 1959) lies in the southwestern corner of the remnant
and shows strong [S II] emission, characteristic of shock-excited
radiative filaments \citep{w66,l83}.  Of great surprise,
however, was the discovery of faint, very thin Balmer-dominated
filaments around almost the entire periphery of the SNR
\citep{l90,s97}.  Such features have been recognized in only a
small number of SNRs and they are interpreted as non-radiative
ionization of partially neutral hydrogen clouds by passage of the
supernova shock, with subsequent recombination to produce the hydrogen
emission lines \citep{c80}.

There has been considerable discussion as to whether G315.4$-$2.3 might
be the remnant of the supernova of AD 185.  Arguments in favor include
the Balmer-dominated filaments and reasonable positional agreement
according to the analysis of historical texts \citep{c77}; arguments 
against include the large size --- a diameter of about
37 pc if it lies at the apparent dynamical distance of 2.8 kpc
\citep{r96} and the possibility that the object observed in 185 may not
have been a supernova at all \citep{c94,s95}. These
arguments were summarized by \citet{s97} who notes that the
distance and whether the SN was of type Ia or II (if it was a SN at
all), are the crucial points of contention.

Improved radio properties of this SNR such as a search for fine scale
features, the detailed magnetic field configuration, and the exact
relation between radio, optical and x-ray features can add more
information to this discussion.  Both the current radio \citep{w96} 
and x-ray \citep{p84,v97} images 
show the same general shell seen in H${\alpha}$ and bright
emission from RCW86 but the radio structure of RCW86 does not match well 
with the different frequency bands.  It has also not been possible to
establish any detailed correspondence of specific individual features
with the limited 45-arcsec resolution of the best radio image available 
from the MOST \citep{w96}.  We have therefore observed
G315.4$-$2.3 with the Australia Telescope Compact Array \citep{a92} 
with a half-power beamwidth of 8 arcsec at a wavelength
of 22 cm to provide higher resolution for both imaging and polarimetry which
previously had only 4.4-arcmin resolution \citep{m74}.  The
equipment and observations will be described in Section 2; the results
will be presented in Section 3; the discussion in Section 4 will
show that this SNR continues to show several unusual properties; and
concluding remarks are in Section 5.

\section{Equipment and Observations}

G315.4$-$2.3 has such a large angular extent that full coverage of the
SNR required a 19-point hexagonal mosaic observation with points spaced 
slightly less that 1/2 the primary beamwidth apart.  Five
configurations of the ATCA were utilized with a total of 70 independent 
baselines covering a range of spacings between 31 and 6000 meters. We note
that much of the overall emission of the SNR is still missing, however.
The observations recover an integrated flux density of approximately 28
Jy, about 70\% of the expected flux density at this frequency
(interpolated from the data of Caswell et al. 1975; Green
1998).  The instrument records the intermediate and fine-scale structure
for comparison with the features seen at the other wavelengths but
does not show the full brightness of the overall shell.

The many pointing shifts during the mosaicing observations caused high
sidelobes with a significant radial spoke pattern.  These can be seen
around a number of point sources and one small-diameter double source were
present in the region of G315.4$-$2.3. In particular, the strong source
with a flux density of 280 mJy inside the remnant toward the southwest
at $\alpha=\rm14^h41^m44.^s48$ and $\delta=-62^\circ34'47.1''$ (all
positions are given for epoch J2000) limited the dynamic range on the
image.  Phase self-calibration was done on
this source for the field centered closest to it and the results were
then transferred to the other pointings with limited success.  At the
position of RCW~86, about 8 arcmin from the point source, we were able
to achieve a dynamic range of about 500/1 which left some ripples in
the image at a level of about 0.6 mJy beam$^{-1}$.  For the five
12-hour observations of the 19-point mosaic pattern, the theoretical
rms noise level per position should have been about 0.03 mJy.  The
ripples can be seen cosmetically in the image but do not affect the 
scientific results.  At the angular distance of the east rim, residual 
effects appear to be less than 0.1 mJy beam{$^{-1}$}, the approximate
rms noise on the final map or about three times the theoretical limit 
from pure receiver noise.  Because the mosaic pattern was large enough to
extend well beyond the SNR, the noise is nearly uniform over the whole 
region shown.   

To avoid phase smearing in the outer regions of the beams and also to
evaluate the Faraday rotation of the polarized emission by utilizing
measurements across the band, the observations were made with a total
of 64 channels each 4 MHz wide in two adjacent IF bands.  After editing
of interference and removal of end channels, a total of 48 channels
with a mean frequency of 1389 MHz was used for the final image.  For
the polarization analysis the data were binned by eight channels into 6
bands with a width of 32 MHz each between 1310 and 1470 MHz.  The six
individual 32-MHz bands were analyzed separately and all six position
angles were used to determine the Faraday rotation measure.  The six
amplitudes were averaged together to produce the final mean polarized
amplitude.  For the vector plots shown below, the amplitudes were
clipped to discard all values below three times the rms noise level of
0.2 mJy beam{$^{-1}$} but for the fractional polarization determination
all polarized intensities down to zero were included.  Individual
position  angle uncertainties were typically $11^{\circ}$ although
values with uncertainties up to $20^{\circ}$ were included in the
analysis.

A MOST image at a frequency of 843 MHz and a resolution of 45 arcsec
was kindly provided by A. Green for comparison purposes.  The
equipment and observing parameters for that image are given in the
paper by Whiteoak and Green (1996).

\section{Results}

\subsection{Total intensity}

The full mosaiced total intensity image is shown in Figure 1a.  The
half-power beamwidth is 8 arcsec.  As well as showing the SNR, this
image illustrates some of the
artifacts present in mosaic images with missing short spacings of the
antennas.  The most striking feature of this supernova remnant
is the lack of any fine-scale features.  It looks virtually
identical to the MOST image, Figure 1b \citep{w96}, with
a resolution of 45 arcsec.  This is not an error in the data analysis
as point sources show the correct 8-arcsec half-power width (compare
Figure 1a with 1b).  Even the edge of
the SNR is not sharp but very diffuse as illustrated by the
representative slices through the shell seen in Figure 2.  The same
effect is even seen in RCW~86 where the emission appears to come
from bright extended patches, not the thin filaments seen in the
optical-line images \citep{s97}.

Among the well-studied remnants, the radio morphology of G315.4$-$2.3
bears the most striking resemblance to Tycho (G120.1+1.4). Both shells
consist of a set of arcs with radii often larger than the average radius
of the periphery. In both, the arcs do not quite close at one location,
leaving a noticable gap, and several of the arcs can be traced into the
interior. G315.4$-$2.3 lacks, however, the sharp outer rims which
delineate much of Tycho's silhouette \citep{dbs91}.

Also reminiscent of Tycho is that the general outline of the circular
shell is similar in both the radio and the
optical (Figure 3).  For the contour display of the radio emission, we
have blanked some of the pixels right around the bright point source at
$\rm14^h41^m44.^s48$ and $-62^\circ34'47.1''$ so as not to distract the
viewer.  The H${\alpha}$ filaments generally lie just
outside the radio shell or just inside it.  The latter placement
could be caused by projection effects of filaments on the shell in
front or back of the plane of the sky.  Projection might also possibly 
explain the very notable lack of radio emission toward the sharp 
filaments on the west side of the remnant which are bright in
H$\alpha$ (Fig.\ 3) and look virtually identical in the x-ray band
\citep{v97}.  These filaments appear to lie
between about .65 and .85 of the radius of the remnant in that
direction so that the projection of a thin shell along the line of
sight would be less than half what it is at the edge.  Such a reduction
in brightness would make their detection marginal in the radio.  

There is no good correlation of individual radio and optical features at
any location in either position, orientation or intensity.  We note that
the east rim has several spots which are nearly as bright as RCW~86 in
the radio but less than 1/50 as bright in H${\alpha}$. Toward RCW~86
itself, the brightest optical emission wraps around the western and
southern sides of the brightest radio patch but lies inside the fainter 
outer radio feature on the southern edge (Fig.\ 4a),
suggesting almost an inverse correlation.  There is, however, a very
faint H$\alpha$ filament just at the outer edge of the radio emission
in that direction. RCW~86 is conclusively
shown to be in radiative equilibrium from the [S II]/H${\alpha}$ ratio
near 1.0 (from the images by \citet{s97}.  

In the north, both the radio and hard x-ray emission \citep{v97} extend
well beyond the optical in several spots.  The optical filament at $\rm
14^h43^m40^s$ and $-62^\circ08'$ is particularly puzzling.  It is
bright in [S~II] as well as H$\alpha$ and is presumably shocked
material in radiative equilibrium but it appears to be sharply broken
at the ends and sits outside a hole of the same length in the main
northern shell.  A radio slice through the region (Fig.\ 2b), on the
other hand, shows that the radio emission just decreases continuously
from the break in the optical shell out to the position of the
apparently broken-off outer piece.  The radio emission then appears to
curve off toward the east and form a bubble which returns to the the
brighter main shell at about $\rm14^h44^m30^s$ and $-62^\circ13'$.
Perhaps the expanding material has hit a small hole in the interstellar
medium.  

In general soft x-rays seem to follow the optical
emission in both the radiative and non-radiative filaments 
but the harder x-rays are different.  The hard/soft x-ray ratio is
generally high in the regions of significant radio emission but there
is certainly not a one-to-one correlation of features.  The general 
configuration was interpreted by \citet{v97} to mean that the 
hard emission represents a non-thermal power law tail to the x-ray 
distribution.

\subsection{Polarized emission}

The polarized emission is too faint to be reliably detected in most
places but can be measured toward RCW~86 and at a few spots on the
eastern rim of the SNR.  The mean fractional polarization within the 2
mJy beam$^{-1}$ total intensity contour toward RCW~86 is 8\% with only
small variations.  The brightest feature on the east limb of 
G315.4$-$2.3 at $\alpha=\rm14^h46^m30^s$ and $\delta=-62^{\circ}33'$,
which has about the
same total intensity as RCW~86, shows no polarization to less than 3\%.
It is only in a few regions further to the north that any reliable
polarization can be measured.  In the most polarized patch at
$\rm14^h45^m05^s$ and $-62^\circ21'$ the polarized fraction reaches about
15\%.  The regions of highest polarized intensity are indeed the same
ones found at 2.7 and 5 GHz with resolutions of $8.'4$ and $4.'4$,
respectively by \citet{m74,m75}. We note that the degree
of polarization is low where the emission of RCW~86 is brightest, and
also where the x-ray emission is most intense. This is consistent with
depolarization caused by differential Faraday rotation in the high
density gas present in those regions.

It is only in the brightest polarized regions that the position angles
of the electric vectors can be measured with sufficient reliability
within the 32-MHz bands to provide an accurate evaluation of the Faraday
rotation.  Using these data we have determined the Faraday rotation and
thence by rotation back to zero wavelength the intrinsic direction of
the {\it magnetic} field.  These results are shown in Figure 5 along with
the fractional polarization.  In order to make them visible, the
magnetic vectors and boxes representing the magnitudes of the Faraday
rotation are plotted at only every tenth pixel for a separation of 2
1/2 beamwidths.  This means that every vector is completely
independent.

The Faraday rotation has a reasonably small range between +20 and +120
rad m$^{-2}$ with most values near +70 rad m$^{-2}$. Typical uncertainties
are around $\pm20$ rad m$^{-2}$. This uncertainty can cause a significant error
(about 50$^\circ$) when rotated back to the intrinsic value at zero
wavelength but the consistency of the adjacent, but independently
measured, vectors would suggest that they appear to be better
determined than their formal errors would indicate.  In the discussion
we shall consider the position angles to be as shown.  

\section{Discussion}

There are several features of G315.4$-$2.3 which make it very reminiscent
of young remnants like Tycho's SNR and SNR$0519-690$ in the LMC.
We see thin Balmer-dominated filaments outside a rather smooth radio
shell with little fine structure.  The filaments are still expanding
with a rather high velocity of 600 km s$^{-1}$ as measured by the
widths of the broad component of the Balmer lines \citep{l90}.
The shell is reasonably complete and circular, although gaps and
irregularities are beginning to appear.  The magnetic field is radially
oriented and the radio spectral index of $-0.61 \pm 0.01$ \citep{r92} is 
somewhat steeper than the average for SNRs. A steeper spectrum is
characteristic of young shell SNRs \citep{d91}.

The presence of forbidden-line emission, primarily in the nebulosity RCW
86, may seem to run counter to the ``youthful" features just mentioned.
However RCW 86 is somewhat reminiscent of the forbidden-line ``fan"
seen in the young SNR Kepler (SN 1604; van den Bergh et al.\ 1973). Even
their shapes are similar, and practically the same spectral lines of
[S~II], [O~II], [O~III], [N~II], etc., are seen in both \citep{l83}. Such 
features are also associated with substantial shock-heated dust, as 
shown by strong infrared emission \citep{g90,b87}.  The proper motions
of their filaments are low, $<$ 0.02 arcsec yr$^{-1}$ or $<$ 260 km
s$^{-1}$ for RCW~86 at the farther distance of 2.8 kpc
\citep{vdb77,k95}; this low a velocity has usually been 
interpreted in terms of swept-up circumstellar material. The one
striking difference is the enhanced x-ray emission around RCW~86.

The nonradiative Balmer line emission, with both narrow and broad
spectral components, is a common feature of young SNRs \citep{s91}, 
though it can also be found in much older objects like the Cygnus
Loop \citep{r83} where, however, radiative forbidden line
emission dominates. From the spectral properties of the Balmer
components, it is possible to derive a shock velocity, though this is
somewhat dependent upon the degree of equilibration \citep{c80}. We 
note that the width of the broad component in the generally
accepted historical shell remnants (Tycho, Kepler and SN1006) ranges
from 1500 to 2300 km s$^{-1}$ \citep{c80,s91}, while in the Cygnus Loop 
a value of about 130 km s$^{-1}$ is
found \citep{h94,r83}. For G315.4$-$2.3, a full-width half-maximum 
of 660 km s$^{-1}$ is observed \citep{l90}. Since the
three historical remnants range in age from 400 --- 1000 yr, while the
Cygnus Loop is usually reckoned to be $\simeq10^4$ years old
\citep{b99}, the intermediate velocity width in G315.4$-$2.3 suggests an
intermediate age of several thousand years. On the low side this could
still be consistent with a supernova as recently as 1800 years ago.

A number of the age indicators mentioned above represent an evolutionary
rather than physical age.
Arguments for a young physical age center on the fact that there may
have been a supernova in about the correct location in AD185 (Clark and
Stephenson 1977; but see Schaefer 1995) and a relatively close distance
of about 1 kpc based on extinction measurements \citep{s94}. The
latter allows the remnant to have a radius of about 6.5 pc so it can
have expanded at a reasonable rate for 1800 years.  Alternatively,
using optical spectroscopy of RCW~86 itself, \citet{r96} found
a mean radial velocity of $-33.2$ km s$^{-1}$ which corresponds to a
distance of 2.8 kpc for the \citet{b93} model of galactic
rotation.  At this distance G315.4$-$2.3 has a radius of about 18 pc.
Expansion to this size in 1800 years implies a mean velocity of $\rm
10\,000\;km\,s^{-1}$. Such a speed could be possible if the remnant was
expanding inside an empty cavity and is just now encountering the inner
walls.  There is a small amount of shock-heated dust in the shell near
RCW86, $< 10^{-3}\rm\, M_\odot$ \citep{g90}, and a ridge of
infrared emission along the eastern edge of the SNR \citep{a89,g90}
which could be from dust in a cavity wall. The
outer walls could still be neutral to give rise to the Balmer-dominated
filaments.  Although the size of the bubble would be too large to have
been produced by the wind of a single progenitor star, \citet{w69} has 
identified an association of O stars surrounding the
position of G315.4$-$2.3 at a distance of about 2.4 kpc.  The supernova
could have occurred in a bubble formed by more than one contributor.

An argument in favor of the larger, kinematic distance comes from mass
estimates.  For an expanding type Ia remnant to have slowed down to 600 km
s$^{-1}$ \citep{l90} from the canonical 10,000 km s$^{-1}$
it must have swept up about 200 times the ejected mass.  Even if it was
a type II supernova with greater ejected mass and slower initial
velocity, the total mass in the SNR must be on the order of 100
$\rm M_\odot$.  

The mass can be determined from the mean density and volume of the SNR.
A number of workers using different techniques from different
wavelength data but a common distance of 1 kpc have arrived at similar
estimates for the density of the material in G315.4$-$2.3.  From an ASCA
pointing covering about 1/5 of the remnant on the northeast side,
\citet{v97} found a mean initial density of about 0.2 cm$^{-3}$, and
$1-10$ cm$^{-3}$ toward RCW86 in the southwest.  \citet{c89} found
0.11 cm$^{-3}$ on the northeast side and 1.1 cm$^{-3}$ toward the
southwest from their EXOSAT data.  \citet{g90} used IRAS
infrared data to get a density of 0.6 cm$^{-3}$ weighted toward the
region around RCW~86.  \citet{s97} has found 0.9 cm$^{-3}$ toward the
Balmer filaments inside the western edge just north of RCW~86.  These
densities are all relatively low and support the idea of a bubble
environment and interaction with a higher density clump on the
southwest.  The difference between northeast and southwest is also
consistent with a density ratio of $\simeq4$ found from theoretical
modeling of G315.4$-$2.3 by \citet{p99}. 

For convenience, we shall assume that 4/5 of the remnant has a mean
density of 0.2 cm$^{-3}$ and 1/5 has 1 cm$^{-3}$.  At 1 kpc the mean
radius is 6.5 pc and the total volume is $3.4\times10^{58}$ cm$^3$. The
total mass is then 6.4 $\rm M_\odot$, well below the necessary amount
to slow the expansion to its present speed.  If the SNR is at 2.8 kpc,
however, the densities, which were determined from the emission
measures, decrease by only the square root of the distance whereas the
volume increases as the cube of the distance.  The resultant mass of 84
$\rm M_\odot$ is certainly within a factor of two of what is required.
Although this argument does place the SNR near 2.8 kpc, it does not
distinguish between a massive progenitor or a low-mass one which just
happened to occur in the vicinity of the association.

At the position of RCW~86, the expansion appears to have encountered a
dense clump of material with sufficiently high density for radiative
equilibrium in the filaments which show a ratio of the bright sulfur
doublet at 671.7 nm and 673.1 nm to H$\alpha$ of about 1.0 \citep{s97}.
Radiative filaments generally show a nearly perfect one-to-one correspondence
between images at radio and optical wavelengths, e.g.\ IC443 \citep{m86},
but in RCW~86 we see that the bulk of the radio emission has no filamentary
structure and lies inside the optical filaments.  The smaller and somewhat
fainter radio clump on the outside of RCW~86 toward the south is apparently
featureless (Fig.\ 3), but looking at the very lowest levels, one sees
that a plateau of faint optical nebulosity actually matches the radio
extension quite well (Fig.\ 4a). The x-ray emission appears
strongest between the optical filaments (Fig.\ 4b) which could be
explained by just-shocked hot gas
which has not yet cooled to radiative equilibrium.  There is possibly
some hardening in the x-ray emission where the radio emission is
brighter \citep{v97}.  The shock may not yet have penetrated all
of the dense cloud and on the inner side, bright radio and x-ray
emission may be enhanced by a reverse shock going back into the ejecta
or previously swept-up material.  The interaction should not extend to
outside the dense feature, however, and we have no explanation for the
peculiar outer radio clump.

The polarization structure observed on this fine scale is similar to
that found by \citet{d76} with a resolution of 8.4 arcmin.
The most highly polarized regions seen with that resolution are also
the brightest in the current study and there is no particular
correlation of polarized intensity and total intensity.  Because of the
greater brightness sensitivity afforded by the lower angular resolution
observations, Dickel and Milne were able to measure the polarization
over the entire SNR and find a nearly radial pattern all around the
shell.  Where measured with the higher resolution on the eastern rim
the new data confirm that pattern of a radially oriented magnetic field
with regard to the center of the SNR.  Toward RCW~86, it starts in a
radial direction but as it approaches the brightest optical part of the
nebula, the field appears to curve off as if it may be wrapping around
the densest part of the cloud.  This appears to be an example of how
irregularities in the interstellar medium can affect the magnetic fields. 

A mean Faraday rotation of $\simeq+60$ rad m$^{-2}$ is found over most
of the remnant. The one significant discrepancy in rotation measure
between the current study and that of \citet{d76} is just on
the eastern side of RCW86 where they found a value of about $-150$ rad
m$^{-2}$. Perhaps in this complex area, convolution over different scale
sizes can create a different rotation. Looking at other polarized
background sources, we find that the $RM$ for two near the
southwest corner: the brightest source which is seen through the
remnant has $RM= +182$ rad m$^{-2}$, while another one (an extended
double) about 12 arcmin to the west of RCW~86 shows values in the range
+25 to +55 rad m$^{-2}$. These values are consistent with the positive
values found for the SNR emission.  In their compendium of rotation
measures, \citet{s81} list nearby sources with both positive and
negative values, although the tendancy appears to be negative. This
agrees with nearby (within $10^\circ$) pulsars, 
three of which have have negative values with a
mean of, $RM=-12$ rad m$^{-2}$, while one (PSR J$1428-55$) has $RM=+4$
rad m$^{-2}$. On the basis of such measurements, Rand and Lyne (1994)
have modelled the interstellar magnetic field, and conclude that in the
general direction of G315.4$-$2.3 the field points away from the solar
system (producing negative $RM$ values). The presence of a galactic
object with $RM>0$ runs counter to their conclusion, suggesting that in
the direction of $l=315^\circ$ the field must have a more complex
structure.

\section{Conclusions}

Although G315.4$-$2.3 has elements indicative of both youth and age, all
of its major characteristics (nearly circular limb-brightened outer shell,
radially-directed magnetic field, spectral index, thin Balmer-dominated
filaments, no detailed correlation between radio and optical emission) are
most frequently seen in young shell SNRs. The radio morphology is most
similar to that of 3C~10 (Tycho), except for the sharp, narrow outer rims
which delineate most of the younger remnant's periphery \citep{dbs91};
there is no trace of such a component in MSH~14$-$6{\it 3}.
It may be significant, however, that the outer rim is not seen along
the eastern edge of 3C~10, where there is evidence of much interaction
with the ambient medium. Perhaps this component, which coincides with
the x-ray edge (and hence the outer blast wave), is a transient
phenomenon which disappears as the blast wave becomes fully adiabatic.
We also note that 3C~10 is unique in showing this distinct outer rim.

RCW~86, the high-density southwestern corner of MSH~14$-$6{\it 3},
remains the most puzzling feature in the remnant, and is perhaps the
key to what is going on. The obvious explanation, that the blast wave
has encountered a high-density cloud, is difficult to rhyme with the
fact that part of this emission extends {\it beyond} the average
radius of the nearly circular shell. An encounter with a dense cloud
should lead to a concave dent in the shell, not to a convex blister.
A completely different possibility is that RCW~86 is where a compact
clump of the exploded star is strongly interacting with the ISM. The
boomerang appearance of the x-ray emission here (Fig.\ 4) strikingly
resembles an outward-moving Mach cone, and similar features have been
identified in the Vela SNR \citep{a95,sjva95}. Such an encounter with 
a dense region would be by chance.

\acknowledgements
We thank Jacco Vink for help with the x-ray data, R. Chris Smith for
providing the optical data, Anne Green for help with the MOST data, and
all three for valuable discussions.  The thoughtful reading of the
manuscript by the referee, Dave Green, is greatly appreciated.  Tom
Brink assisted with some of the early data analysis.  JRD gratefully
acknowledges a Visitor's Fellowship from the Netherlands Foundation for
Scientific Research (NWO) during his very enjoyable stay at ASTRON.

\clearpage



\figcaption[f1a.eps,f1b.eps]{a) A greyscale image of the supernova remnant
  G315.4$-$2.3 and surroundings at a frequency of 1.34 GHz.  The scale
  for the wedge is Jy beam$^{-1}$. The 8-arcsec beam is shown as
  the tiny dot in the lower right corner of the plot.  b)  A MOST image
  with a resolution of 45 arcsec shown for comparison.  The range of
  the greyscale is from 0 to 300 mJy beam$^{-1}$.  Artifacts including
  the 1.2$^{\circ}$ grating ring are discussed by \citet{w96}. \label{fig1}}

\figcaption[f2a.eps,f2b.eps,f2c.eps,f2d.eps,f2e.eps]{Sample
  slices through the radio image of G315.4$-$2.3 --- a) through the
  eastern rim;  b) through the northern ``breakout'' region c)
  through the northern rim and including a faint point source; and d)
  through the region of RCW~86 itself.  All the plots are to an
  identical scale. \label{fig2}}

\figcaption[f3.eps]{H$\alpha$ greyscale (Smith 1997) and selected 1.34
  GHz radio contours of the supernova remnant G315.4$-$2.3. \label{fig3}}

\figcaption[f4a.eps,f4b.eps]{The region of RCW~86.  a) A
  logarithmic H$\alpha$ greyscale image in an attempt to show both the
  faint and bright emission with superimposed selected radio
  contours.  For the contour display, the radio data have been blanked within
  4 arcmin of the bright point source which would lie just at the top
  of the wedge on the northeastern edge of the image.  b) A linear
  H$\alpha$ greyscale over a limited range with selected x-ray contours
  from the public ROSAT HRI image. \label{fig4}}

\figcaption[f5a.eps,f5b.eps]{Radio polarimetric data on
  G315.4$-$2.3 at 1.34 GHz.  The contours are the total intensity with
  values of 2, 4, and 6 mJy beam$^{-1}$.  The greyscale is the
  fractional polarization with the ranges indicated on the wedges.  The
  boxes indicate the Faraday rotation measure.  The vectors show the
  direction of the $\it magnetic$ field with the length 
  proportional to the polarized intensity.  a) The east rim of the SNR.  
  Only a few faint boxes and vectors are present near the northern end
  and two toward the center of the plot.  A box diameter of 10 arcsec
  represents a rotation measure of 38 rad m$^{-2}$ and a vector length
  of 10 arsces is a polarized flux density of 0.23 mJy beam$^{-1}$  b)
  The RCW~86 region.   A box diameter of 10 arcsec represents a
  rotation measure of 58 rad m$^{-2}$ and a vector length of 10 arcsec
  is a polarized flux density of 0.57 mJy beam$^{-1}$. \label{fig5}} 








\end{document}